\documentclass[a4paper,11pt]{JHEP3}
\usepackage{fancyheadings}
\usepackage[dvips]{graphicx}
\usepackage{ifthen}
\usepackage{amsmath}
\usepackage{epsfig}
\usepackage[errorshow]{tracefnt}
\newcommand\nn{\nonumber}

\newcommand\be{\begin{equation}}
\newcommand\beal{\begin{align}}
\newcommand\eeal{\end{align}}
\newcommand\benu{\begin{enumerate}}
\newcommand\eenu{\end{enumerate}}
\newcommand\bit{\begin{itemize}}
\newcommand\eit{\end{itemize}}

\newcommand{\ee}{\end{equation}}
\newcommand{\bd}{\begin{displaymath}}
\newcommand{\ed}{\end{displaymath}}

\newcommand\ga{\gamma}

\newcommand\wps{{\widehat{\Psi}}}
\newcommand\psp{{\Psi^+}}
\newcommand\psm{{\Psi^-}}

\title{M-theory on eight-manifolds revisited: \\
${\cal N}=1$ supersymmetry and generalized $Spin(7)$ structures}

\author{Dimitrios Tsimpis \\
Max-Planck-Institut f\"{u}r Physik --Theorie\\
F\"{o}hringer Ring 6,  80805 M\"{u}nchen, Germany\\
E-mail: \email{tsimpis@mppmu.mpg.de}
}

\abstract{The requirement of ${\cal N}=1$ supersymmetry 
for M-theory backgrounds of the form of a warped product 
${\cal M}\times_{w}X$,  
where $X$ is an eight-manifold 
and ${\cal M}$ is 
three-dimensional Minkowski or AdS space, 
implies the existence of a 
nowhere-vanishing 
Majorana spinor $\xi$ on $X$. 
$\xi$ lifts to a nowhere-vanishing spinor on the 
auxiliary nine-manifold $Y:=X\times S^1$, where 
$S^1$ is a circle of constant radius,  
implying the reduction of the structure group of $Y$ 
to $Spin(7)$. In general, however, there is no 
reduction of the structure group of $X$ 
itself. 
This situation can be described in the language of 
generalized $Spin(7)$ structures, defined in terms of certain 
spinors 
of $Spin(TY\oplus T^*Y)$. We express the 
condition for ${\cal N}=1$ supersymmetry in terms of 
differential equations for these spinors. 
In an equivalent formulation, working locally 
in the vicinity of any point in $X$ 
in terms of a `preferred' $Spin(7)$ structure, 
we show that the requirement of 
${\cal N}=1$ supersymmetry amounts to 
solving for the intrinsic torsion 
and all  
irreducible flux components, except for the one lying in the 
$\bf{27}$ of $Spin(7)$, in terms of the warp factor and a one-form 
$L$ on $X$ 
(not necessarily nowhere-vanishing) constructed as a $\xi$ bilinear; in addition, 
$L$ is constrained to satisfy a pair of differential equations. 
The formalism based on the group $Spin(7)$ 
is the most suitable language in which 
to describe supersymmetric compactifications on eight-manifolds of 
$Spin(7)$ structure, and/or small-flux perturbations 
around supersymmetric compactifications on manifolds of $Spin(7)$ 
holonomy. 
}

\preprint{MPP-2005-129}
\begin{document}

\section{Introduction}

It has been observed (starting with \cite{gaun}), 
in connection to supergravity compactifications, that the concept 
of $G$-structures is a natural generalization of 
special-holonomy to the case where fluxes are present. Supersymmetry 
implies the existence of a nowhere-vanishing spinor 
on the internal 
manifold $X$, thereby reducing the structure group of $X$ 
to $G$. 
In the presence of fluxes, 
$X$ is no longer 
special-holonomy and the spinor is no longer covariantly constant:  
its failure to be such is parametrized by the (flux-dependent) intrinsic torsion 
of the Levi-Civita connection associated with the $G$-invariant metric on $X$. 
Moreover, the intrinsic torsion can be decomposed in irreducible $G$-modules, giving 
a characterization of $X$.

More recently, it was realized 
\cite{gg3, gg4, jescb, grana, granb, jesca} that 
generic spinor Ans\"{a}tze for the supersymmetry parameter 
naturally lead to the concept of generalized $G$-structures \cite{hitchin}. 
Roughly-speaking,  generalized $G$-structures arise as follows: 
typically there will be two nowhere-vanishing spinors $\epsilon^{\pm}$ in 
 the Ansatz for the supersymmetry parameter, each one 
inducing a reduction of the structure group to a subgroup $G_{\pm}$.  
Noting, in addition, that there is an isomorphism 
between bispinors on $X$ and spinors of $TX\oplus T^*X$, 
we conclude that $\rho:=\epsilon^+\otimes\epsilon^-$ 
(which can also be thought of --by Fierzing-- 
as a sum of forms on $X$) induces a reduction 
of the structure group  of $TX\oplus T^*X$ to $G_{+}\times G_-\subset Spin(TX\oplus T^*X)$.

In addition to the reduction of the structure group of $TX\oplus T^*X$, 
supersymmetry implies that $\rho$ should satisfy certain 
differential equations. For type II supergravities 
and for $X$ a six- or seven-dimensional 
manifold, these equations have been identified, in the case where the 
Ramond-Ramond fields are zero,  with certain integrability conditions 
for the generalized structures \cite{gg3, jescb}. Recently 
it has been possible to 
give a  satisfactory mathematical description 
of the RR forms \cite{jesca} by a generalization of the Hitchin functional \cite{hitb} 
in which the RR forms appear as constraints. 
The role of the Hitchin functional 
in various topological theories is explored in 
\cite{gera, dijk, nekr, angu, pest}. 

Although a great deal is known about the connection of 
supersymmetry to generalized structures in 
six and seven dimensions, the case of eight-dimensional manifolds 
remains rather obscure (see however \cite{1,2}). 
In the present paper we wish to remedy 
the situation by examining the conditions 
for the most general ${\cal N}=1$ three-dimensional AdS or Minkowski 
vacua in M-theory. An immediate consequence of 
supersymmetry is that associated with the 
eight-dimensional internal manifold  $X$ 
there is a nine-manifold $Y:=X\times S^1$, such that 
$Y$ supports a  generalized Spin(7) structure on the sum of its tangent 
and cotangent bundles. The structure is given in terms of 
certain bispinors (spinors of $Spin(9,9)$) which are constrained 
to satisfy certain differential equations.

In a more conventional (equivalent) formulation, we show that 
${\cal N}=1$ supersymmetry implies 
the existence of a nowhere-vanishing {\it Majorana} spinor 
on $X$. This lifts to 
a  nowhere-vanishing spinor on $Y=X\times S^1$ 
and hence implies the reduction of the structure group 
of $Y$ to $Spin(7)$. 
Note that, in general, $X$ does {\it not} support 
nowhere-vanishing {\it Majorana-Weyl} spinors and the 
structure group of $X$ itself is {\it not}, in general, reduced. 
However, 
working locally in an open set of $X$, one can still 
decompose all fields 
in terms of irreducible $Spin(7)$-modules. We are then able to 
solve for the intrinsic torsion and all  
irreducible flux components, except for the one lying in the 
$\bf{27}$ of $Spin(7)$ which 
cannot be determined by the supersymmetry equations, 
in terms of the warp factor and a certain one-form 
$L$ on $X$ (constructed as a $\xi$ bilinear).  
$L$ is not necessarily nowhere-vanishing, 
unless the structure group is further reduced to $G_2$. 
In addition, 
$L$ is constrained to satisfy a pair of differential equations.

The case examined here is an 
alternative formulation to the works of  
\cite{ba, bb}, which assume the existence of 
at least one nowhere-vanishing Majorana-Weyl spinor, as well as
 of \cite{ms} (see also \cite{bc}). The authors of 
reference \cite{ms} perform their analysis 
by decomposing the flux and the intrinsic torsion of the 
internal manifold in terms of 
irreducible $G_2$ representations. These 
decompositions are valid  only outside the zero locus  
of both Majorana-Weyl spinors $\xi^{\pm}$, where 
$\xi=\xi^+\oplus\xi^-$, and  they break down at the points 
where either of $\xi^{\pm}$ vanishes. The analysis 
of \cite{ms} is valid globally only if there is a further reduction of the 
structure group of $X$ to $G_2$. Note, however, that in open 
sets where neither of $\xi^{\pm}$ vanishes, the analyis 
of \cite{ms} is perfectly 
sufficient to determine the most general local form of the 
geometry. In such opens sets, our formul{\ae} in section \ref{analysis}  
below should reduce to the corresponding formul{\ae} given in \cite{ms}, to the 
extent they overlap (some of the flux components were 
not given explicitly in \cite{ms}).

There are certain cases in which it is advantageous to work with a 
$Spin(7)$ rather than with a $G_2$ structure. Clearly this is true 
if one 
wishes to consider supersymmetric 
M-theory compactifications on eight-manifolds with a 
global $Spin(7)$ structure. A (very) special case thereof is compactifications 
on eight-manifolds with $Spin(7)$ holonomy. Also, even in the generic case 
of compactifications on eight-manifolds $X$ where there is no reduction of the 
structure group of $X$ to $Spin(7)$, it is still 
advantageous to work with a  
$Spin(7)$ strucutre if one wishes to describe the 
local geometry in the vicinity of a point where either of $\xi^{\pm}$ vanishes. 
Put in another way: let $P$ be an arbitrary point in $X$. There is always an open 
set $U_P\subset X$ such that $P\in U_P$ and at 
least one of $\xi^{\pm}$ is nowhere-vanishing 
in $U_P$. I.e. for any point $P$, 
there is always an open patch $U_P$ containing $P$, such that 
the decomposition in $Spin(7)$ modules is valid in $U_P$. The same is {\it not} 
true for $G_2$, as can be seen by taking $P$ to be a point where 
one of $\xi^{\pm}$ vanishes. 

The plan of this paper is as follows: section \ref{gstr} 
includes some general background on $G$-structures 
and, more particularly, $Spin(7)$-structures. 
In section \ref{nsusy} we perform the   
supersymmetry analysis in terms of  (ordinary) 
$Spin(7)$-structures on $X$. As an 
explicit example we have  also included a 
small-flux perturbation around the special-holonomy 
solution involving the non-compact $Spin(7)$-manifold of 
\cite{sala, gibb}. Section \ref{generalized} includes the supersymmetry 
analysis in terms of generalized $Spin(7)$ structures.  The final section 
includes some discussion of future directions. To 
improve the presentation of the paper, almost all of the 
technical details of the supersymmetry analysis have been relegated 
to the appendices.

\section{G-structures and intrinsic torsion}
\label{gstr}

In this section we give a brief review of 
G-structures and intrinsic torsion with emphasis on the 
points relevant to our case.  

Quite generally, the requirement of supersymmetry 
for backgrounds of the form ${\cal M}\times_{w} X$, 
where ${\cal M}$ is maximally-symmetric, implies the existence 
of a nowhere-vanishing spinor $\xi$ satisfying 
a Killing equation
\beal
\nabla_m\xi=G_m\xi~,
\label{intr}
\end{align}
where $\nabla_m$ is the Levi-Civita connection and 
$G_m$ is a vector field on $X$ taking values in the  Clifford algebra 
Cliff($d$), ~$d:={\rm dim}_{\mathbb{R}}(X)$. 
The exact expression depending on the supergravity under consideration, $G_m$ is determined 
by the fluxes  (i.e. antisymmetric tensor fields)  
on $X$ and is generally nonzero. If $G_m$ vanishes identically, $X$ is a 
special-holonomy manifold.

Typically, the existence of $\xi$ implies the 
reduction of the structure group to 
a subgroup $G\subset Spin(d)$ and the manifold 
$X$ admits a $G$-structure. The  latter is characterized by the 
intrinsic torsion, which is a measure of the failure of $\xi$ 
to be covariantly constant with respect to the 
connection associated with the metric induced by the 
$G$-structure. From what has been just said, it is clear that 
the intrinsic torsion could be read off of $G_m$
 in equation (\ref{intr}) 
above. Hence 
the result that the intrinsic torsion 
can be expressed in terms of the fluxes. Schematically:
\beal
{\rm flux}\longrightarrow G_m\longrightarrow {\rm intrinsic ~torsion}~.
\end{align}
Typically, the Killing spinor $\xi$ gives rise to certain 
$G$-invariant forms ($\Phi$), constructed out of $\xi$ 
as spinor bilinears. It is an important 
result that the intrinsic torsion ($\omega$) can also be 
read off of the 
exterior derivatives of these forms:
\beal
\nabla\xi\longleftrightarrow \omega \longleftrightarrow d\Phi~.
\label{sce}
\end{align}
Having a dictionary  between these two alternative 
descriptions is very useful in practice, as it 
allows us to express the supersymmetry equation (\ref{intr}) 
in purely algebraic form. We will see how this works explicitly in section 
\ref{analysis}.

The intrinsic torsion  can be decomposed in terms of 
irreducible $G$-modules: $\omega\in \Lambda^1\otimes \mathfrak{ g}^{\perp}$, 
where $\mathfrak{ g}^{\perp}$ is the complement of $\mathfrak{ g}:={\rm Lie}(G)$ inside 
$spin(d)$. Special classes of manifolds arise when some of these modules vanish; for example 
when all the modules vanish the manifold is special-holonomy. The construction/classification
of manifolds according to their intrinsic torsion is a 
difficult problem which still remains largely open. 
There is of course great interest from the physics point-of-view because of the 
connection to flux compactifications. In some cases it is not known if examples of manifolds 
exist in all classes of possible 
combinations of nonzero modules.


\subsection{Spin(7) structure}
\label{spinstructure}

Let us now see how the general discussion of $G$-structures applies to the case 
where $X$ is an eight-manifold. The main result  
explained here is that the existence of a nowhere-vanishing Majorana spinor 
 on $X$,  $\xi=\xi^+\oplus\xi^-$ 
where $\xi^+$ ($\xi^-$) is of positive (negative) chirality, 
induces the reduction of the structure group of 
the associated nine-manifold $Y:=X\times S^1$, 
where $S^1$ is a circle of 
constant radius, to $Spin(7)$.

A $Spin(7)$ structure on $X$ is a principal sub-bundle of the 
frame bundle over $X$, with fiber the subgroup $Spin(7)$ of $GL(8,\mathbb{R})$. 
We can give alternative description of the $Spin(7)$ structure as follows \cite{joyce}: 
let $x^1,\dots ,x^8$ be the coordinates of $\mathbb{R}^8$. The self-dual four-form
\beal
\Phi^+_{0}&:=e^{1234}+e^{1256}+e^{1278}+e^{1357}-e^{1368}-e^{1458}-e^{1467}\nn\\
&-e^{2358}-e^{2367}-e^{2457}+e^{2468}+e^{3456}+e^{3478}+e^{5678}~,
\end{align}
where $e^{ijkl}$ denotes $dx^i\wedge dx^{j}\wedge dx^{k}\wedge dx^l$, is fixed 
by $Spin(7)\subset GL(8,\mathbb{R})$. At each point $p\in X$ let us define 
$\mathcal{A}_pX$ to be the subset of four-forms $\Phi_p\in \Lambda^4T^*_pX$ for which 
there exists an isomorphism between $T_pX$ and $\mathbb{R}^8$ such that 
$\Phi_p$ is identified with $\Phi^+_{0}$. It follows that $\mathcal{A}_pX$ 
is isomorphic to $GL(8,\mathbb{R})/Spin(7)$. Let $\mathcal{A}X$ be the bundle over $X$ 
with fiber $\mathcal{A}_pX$ for each $p\in X$. We say that a four-form $\Phi$ on $X$ 
is {\it admissible} if $\Phi_p\in \mathcal{A}_pX$ for each $p\in X$. In other words,  
admissible forms are those that can be `reached' from $\Phi^+_{0}$, at each point in $X$. 
It follows that there is a 1-1 correspondence between $Spin(7)$ structures and 
admissible four-forms $\Phi\in \mathcal{A}X$.

The isotropy group of a nonzero Majorana-Weyl spinor of $Spin(8)$,  
is $Spin(7)$. This simply follows from the fact that 
under $Spin(7)\subset Spin(8)$ the chiral spinor representation of $Spin(8)$ 
decomposes as $\bf{8}\rightarrow \bf{7}+\bf{1}$, i.e.
 there is a singlet in the decomposition. Hence a 
nowhere-vanishing Majorana-Weyl spinor of $Spin(8)$ induces a reduction of the 
structure group of $X$ to $Spin(7)$. An equivalent 
way to understand the reduction of the structure group is by noting that 
there is a nowhere-vanishing self-dual four-form 
which can be constructed as a bilinear of the chiral spinor. 


Let $I_{\pm}$ be the isotropy groups of $\xi^{\pm}$. 
The isotropy 
group $I_+\cap I_-$ of $\xi=\xi^+\oplus \xi^-$ induces a 
local reduction of the structure group. 
If both Majorana-Weyl spinors $\xi^{\pm}$ are nowhere-vanishing, the structure group 
of $X$ reduces to the common subgroup of the two $Spin(7)$ structures: 
$Spin(7)_+\cap  Spin(7)_-=G_2$. The reduction 
of the structure group to $G_2$ can be deduced alternatively from the fact 
that in this case there exist  
a nowhere-vanishing vector and a three-form on $X$, which can be constructed 
as a bilinears of $\xi^{+}$ and $\xi^{-}$.

In general, both chiral spinors may have zeros. At a point where $\xi^{+}$  ($\xi^{-}$) 
vanishes,  $I_{+}$ ( $I_{-}$) is enhanced to $Spin(8)$ 
and the isotropy group of $\xi=\xi^+\oplus\xi^-$ is enhanced to 
$I_+\cap I_-=Spin(7)_{\mp}$. 
This, however, does not induce a reduction of 
the structure group to $Spin(7)$ unless, as explained in the 
previous paragraph, $X$ supports a nowhere-vanishing 
Majorana-Weyl spinor. Roughly-speaking, 
the isotropy group of $\xi$ is 
not, in general, a fixed $Spin(7)$ subgroup of $Spin(8)$ -- 
 as is required for a reduction of the structure group: 
as one moves around in $X$,  
$I_+\cap I_-$ `rotates' inside $Spin(8)$.

Nevertheless, it is possible to translate this situation to an 
honest $Spin(7)$ reduction of the structure group not of $X$ itself, but 
of an associated nine-manifold. 
First, let us recall some useful facts about $Pin$ vs $Spin$ groups.  
The reader can consult, e.g., \cite{spin} for more details. 
The group $Pin(n)$ sits inside ${\rm Cliff}(n)$, therefore 
the irreducible representations of ${\rm Cliff}(n)$ restrict to representations 
of $Pin(n)$ -- which actually turn out to be irreducible as well. In 
particular, the real irreducible 
representation of $Pin(8)$ is the restriction of the real irreducible 
representation of ${\rm Cliff}(8)$ -- the 
sixteen-dimensional Majorana spinor in eight dimensions. 
Similarly, since $Spin(n)$ sits inside ${\rm Cliff}_0(n)$ (the even part of the 
Clifford algebra), the irredicible representations 
of ${\rm Cliff}_0(n)$ restrict to representations of $Spin(n)$ -- 
which also turn out to be irreducible. Furthermore, there is an important 
isomorphism:
\beal
{\rm Cliff}(n) \cong {\rm Cliff}_0(n+1)  ~.
\label{spiniso}
\end{align}
Coming back to physics: we are considering M-theory 
compactifications on an eight-manifold $X$. 
Supersymmetry imposes the existence of a nowhere-vanishing real 
pinor $\xi$ on $X$. Tensoring $X$ by a 
constant circle $S^1$, we obtain 
a compact nine-dimensional product 
manifold $Y:=X\times S^1$. 
As can be seen from (\ref{spiniso}), the nowhere-vanishing pinor $\xi$ on $X$ 
lifts to a nowhere-vanishing 
spinor on $Y$ in the 
real, sixteen-dimensional irreducible representation of 
$Spin(9)$. Of course $\xi$, thought of as a spinor on $Y$,  
does not depend on the co-ordinate of $S^1$. It is well-known that 
the existence 
of a nowhere-vanishing spinor on a nine-manifold 
$Y$ induces the reduction of 
the structure group of $Y$ to $Spin(7)$\footnote{Recall that  
$Spin(9)$ acts transitively on the 
unit sphere in sixteen dimensions, $S^{15}\cong Spin(9)/Spin(7)$, 
and that a real, sixteen-dimensional 
unit spinor on $Y$ implies the existence of a global section 
of the sphere bundle over $Y$ with fiber $S^{15}$.}.

Alternatively, the reduction of the structure group of $Y$ 
can be understood as follows: 
Since $\xi$ is nowhere-vanishing, the Majorana spinors 
$\epsilon^{\pm}:=\frac{1}{\sqrt{2}}(\xi^+\pm \xi^-)$ 
(the normalization is for later convenience) 
are also nowhere-vanishing and each of them induces a reduction of the 
structure group of $Y$ to $Spin(7)$. At the points in $Y$ where 
$\epsilon^{\pm}$ are not parallel, the isotropy group 
of $\epsilon^{\pm}$ is reduced to 
$G_2=Spin(7)\cap  Spin(7)$. However, $\epsilon^+$ becomes 
parallel to $\epsilon^-$ precisely at the points where either 
of $\xi^{\pm}$ vanishes. At these points the isotropy group is enhanced
to $Spin(7)$. 
This point-of-view is better suited for the description in terms of 
{\it generalized} $Spin(7)$ structures, which we introduce in section 
\ref{generalizeds} below.

The topological obstruction to the 
existence of a nowhere-vanishing Majorana-Weyl spinor on $X$ is known 
\cite{gray, isham}: it is equivalent 
to the condition that the Euler number of $X$ be 
given by 
\beal
\chi(X)=\pm\frac{1}{2}\int_{X}(p_2-\frac{1}{4}p_1^2)~,
\label{top}
\end{align}
where $p_{1,2}$ are the first and second Pontrjagin forms. 
The sign 
on the right-hand side 
of the equation above depends on the chirality of the nowhere-vanishing spinor. 
As is it follows immediately 
from (\ref{top}), 
requiring that nowhere-vanishing Majorana-Weyl spinors of {\it both} chiralities exist,   
leads to the condition that the Euler number of $X$ vanishes. This, of course, 
is exactly the topological condition for the existence of 
a nowhere-vanishing vector field on $X$. 
Indeed, a vector field can be constructed 
as a bilinear of $\xi^{+}$ and $\xi^{-}$ and, interestingly, equations (\ref{top}) 
can be related to the condition $\chi(X)=0$ by 
a certain triality rotation \cite{isham}. As we have seen, in this case 
there is a reduction of the structure group to $G_2$. In the generic case, 
although there is a 
nowhere-vanishing Majorana spinor $\xi=\xi^+\oplus\xi^-$ on $X$, 
both $\xi^{\pm}$ may have zeros; hence  $X$ need {\it not} 
satisfy equation (\ref{top}).

\section{${\cal N}=1$ supersymmetry}
\label{nsusy}

The starting point 
of our analysis is the supersymmetry equations given in 
(\ref{1}-\ref{4}) below. We will now 
 describe the Ansz\"{a}tze leading 
up to these equations, as well as some basic background 
on eleven-dimensional supergravity in order to establish  
conventions. This brief review follows \cite{ms}.

The field content of eleven-dimensional supergravity 
consists of a metric, a Majorana vector-spinor (gravitino) 
and a four-form field strength $G$. We shall consider eleven-dimensional M-theory 
backgrounds of the form 
of a warped product ${\cal M}\times_{w}X$, 
where $X$ is an eight-manifold 
and ${\cal M}$ is 
three-dimensional Minkowski or AdS space. Explicitly, 
the metric Ansatz reads
\beal
ds_{11}^2=e^{2\Delta}(ds_3^2+g_{mn}dx^mdx^n)~,
\end{align}
where $e^{2\Delta}$ is the warp factor and 
$ds_3^2$ is the metric on $\mathcal{M}$. For the convenience 
of the reader, we follow the notation of \cite{ms}. 
The most general four-form flux Ansatz respecting three-dimensional 
covariance reads
\beal
G=e^{3\Delta}(F+Vol_3\wedge f)~,
\end{align}
where $Vol_3$ is the volume element along the noncompact directions and 
$f$ ($F$) is a one-form (four-form) on $X$. 
Finally, the eleven-dimensional supersymmetry parameter $\zeta$ splits into 
a direct product of a Majorana spinor $\psi$ on ${\cal M}$ and 
a spinor $\xi=\xi^+\oplus\xi^-$ on $X$: 
\beal
\zeta=e^{-\frac{\Delta}{2}}\psi\otimes(\xi^+\oplus\xi^-) ~.
\end{align}
More precisely, $\xi$ is a section of the real spin sub-bundle 
$S_{\mathbb{R}}^{+}\oplus S_{\mathbb{R}}^{-}$ on $X$, where 
$S^{+} \oplus S^{-}$ is the spin bundle on $X$ and 
$S^{\pm}=S_{\mathbb{R}}^{\pm}\otimes\mathbb{C}$. 
Furthermore, since ${\cal M}$ is Minkowski or AdS, $\psi$ is constrained to 
satisfy
\beal
\nabla_{\mu}\psi+m\ga_{\mu}\psi=0~,
\end{align}
where 
$m$ is a massive parameter proportional to the 
inverse radius of ${\cal M}$. 
Substituting our Ans\"{a}tze into the supersymmetry transformations 
of eleven-dimensional supergravity, we arrive at the following equations.

Internal gravitino:
\beal
0&=\nabla_m\xi^++\frac{1}{24}F_{mpqr}\ga^{pqr}\xi^--\frac{1}{4}f_n\ga^n{}_m\xi^+-m\ga_m\xi^-
\label{1}\\
0&=\nabla_m\xi^-+\frac{1}{24}F_{mpqr}\ga^{pqr}\xi^++\frac{1}{4}f_n\ga^n{}_m\xi^-+m\ga_m\xi^+
\label{2}~.
\end{align}
External gravitino:
\beal
0&=\frac{1}{2}\ga^m\partial_m\Delta\xi^+
-\frac{1}{288}F_{mpqr}\ga^{mpqr}\xi^--\frac{1}{6}\ga^n{}f_n\xi^++m\xi^-
\label{3}
\\
0&=\frac{1}{2}\ga^m\partial_m\Delta\xi^-
-\frac{1}{288}F_{mpqr}\ga^{mpqr}\xi^++\frac{1}{6}\ga^n{}f_n\xi^--m\xi^+
\label{4}~.
\end{align}
We have thus rewritten the 
eleven-dimensional supersymmetry transformations purely 
in terms of fields on $X$.

In addition to the supersymmetry equations, a solution of eleven-dimensional supergravity 
should satisfy the Bianchi identities and the equations-of-motion. 
It can be shown that these take the form\footnote{
The analysis of \cite{ms} suggests that the last line in 
(\ref{eom}) may be redundant, i.e. it may follow from the 
supersymmetry equations and the remaining equations in (\ref{eom}). We 
thank D.~Martelli for discussions on this point.}
\beal
0&=d(e^{3\Delta} F)\nn\\
0&=e^{-6\Delta}d(e^{6\Delta}\star f)-\frac{1}{2}F\wedge F\nn\\
0&=e^{-6\Delta}d(e^{6\Delta}\star F)-f\wedge F~.
\label{eom}
\end{align}
One can show that, under a certain mild condition which is satisfied 
for the backgrounds considered in this paper, the supersymmetry equations 
together with the Bianchi identities and equations-of-motion (\ref{eom}) 
imply the Einstein equations \cite{inta}. 
Similar integrability statements also hold for IIA \cite{intb} 
and IIB \cite{intc} supergravities.

Note that setting $m=0$ excludes any solutions with nonzero fluxes 
if $X$ is smooth and compact. As noted in \cite{ms}, 
this can be seen immediately from the scalar part of the Einstein equation:
\beal
e^{-9\Delta}\Box e^{9\Delta}-\frac{3}{2}|F|^2-3|f|^2+72m^2=0~.
\end{align}
Integrating by parts gives $f$, $F=0$. 
This no-go `theorem' can be evaded by allowing the equations-of-motion and/or 
Bianchi identities to be modified, e.g. by introducing source terms or 
higher-order curvature corrections.  

A well-known higher-order correction is the one related to the M five-brane 
anomaly \cite{mina}
\beal
d\star G+\frac{1}{2}G\wedge G=\beta X_8~,
\label{corr}
\end{align}
where $\beta$ is a constant of order $l_{Planck}^6$ and 
$X_8$ is proportional to the same combination of Pontrjagin forms appearing on 
the right-hand side of (\ref{top}). However, generally it is {\it inconsistent} 
to only include the correction (\ref{corr}) without considering 
the corresponding order-$l_{Planck}^6$ corrections 
to the supersymmetry equations. The latter corrections 
are, unfortunately, unknown to date\footnote{In certain cases, e.g. compactification 
on `large' eight-manifolds, it is in fact consistent to ignore all higher-order corrections 
except for the one in (\ref{corr}), see \cite{peet} for a detailed argument.}.

The implications of the supersymmetry equations above are examined 
in section \ref{generalizeds} 
from the point of view of 
generalized Spin(7) {}structures. 
The more conventional  approach is pursued in section \ref{analysis}. 
Before we close this section, let us make an observation which will 
be important in the following: 
as we explain in appendix \ref{susyd}, it follows from (\ref{1}, \ref{2}) that 
the Majorana spinor $\xi$ has constant 
norm,  which we can normalize to unity without loss of generality:
\beal
|\xi|^2=|\xi^+|^2+|\xi^-|^2=1~.
\label{x}
\end{align}
This equation was first noticed in \cite{ms}.

\subsection{Analysis}
\label{analysis}

It follows from equation (\ref{x}) that at each point $p$ in $X$ at least one of 
$\xi^{\pm}$, let us say $\xi^+$ for concreteness,  is non-vanishing. 
In an open set around $p$, we can parameterize:
\beal
\xi^+&=\frac{1}{\sqrt{1+L^2}}\eta\nn\\
\xi^-&= \frac{L_m}{\sqrt{1+L^2}}\ga^m \eta~,
\end{align}
where $\eta$ has unit norm: $|\eta|^2=1$. 
Note that the one-form $L$ can be thought of as 
a $\xi^+$, $\xi^-$ bilinear. 
Moreover, 
we can define a self-dual four-form $\Phi$ as an $\eta$-bilinear via
\footnote{In our conventions the 
real spinor $\eta$ satisfies $\eta^{\dagger}=\eta^{Tr}C$, where 
$C$ is the charge-conjugation matrix. We use $C$ to raise/lower indices 
on the gamma-matrices, so that the notation $\eta\gamma_m\eta$ is 
a shorthand for $\eta^{Tr}(C\gamma_m)\eta$, etc. }
\beal
\Phi_{mpqr}:=\eta\ga_{mpqr}\eta~.
\label{s7}
\end{align}
As was explained in detail in section \ref{spinstructure}, 
the existence of the 
four-form defined in (\ref{s7}) above induces a 
$Spin(7)$-structure on $X$; therefore, one can decompose all 
fields in terms of irreducible representations of $Spin(7)$. 
As was  mentioned in section \ref{gstr}, 
it is very useful to be able to translate back and forth between the spinor and the G-structure 
language; 
in this way the supersymmetry equations can be expressed as a set of purely algebraic 
relations. For the case at hand, the schematic equation (\ref{sce}) is 
nothing but the 
statement that the following two equations are equivalent:
\begin{center}
\fbox{\parbox{8.7cm}{
%
%
\beal
\partial_{[m}\Phi_{pqrs]}&=-8\Phi_{[mpqr}\omega^1_{s]}
-\frac{4}{15}\varepsilon_{mpqrs}{}^{ijk}\omega^2_{ijk}\nn\\
\nabla_m\eta&=\Big\{\omega^1_n\ga^n{}_m +\omega^2_{mpq}\ga^{pq}\Big\}\eta\nn~,
\end{align}
%
}}
\end{center}
\be\label{a}\end{equation}
where $\omega^1$ transforms in the $\bf{8}$ of $Spin(7)$ while $\omega^2$ 
transforms in the $\bf{48}$. 
Note that $d\Phi$ being a five-form 
it transforms in the $\bf{8}\oplus\bf{48}$ of $Spin(7)$, 
hence the decomposition on the right-hand side of the first line 
of (\ref{a}).  $\omega^{1,2}$ generate the two modules of 
the intrinsic torsion of 
a manifold of $Spin(7)$ structure \cite{fern}. 
The equivalence of the two equations in (\ref{a}), is proven in appendix \ref{equivalence}.

Skipping all the details of the derivation, which can be found 
in appendix \ref{susyd}, the supersymmetry conditions are equivalent to the following equations:  
the one-form $L$ is constrained to satisfy

\begin{center}
\fbox{\parbox{8cm}{
%
%
\beal
d\Big(e^{3\Delta}
\frac{L}{1+L^2}
\Big)&=0\nn\\
e^{-12\Delta}\star d\star \Big(
e^{12\Delta}\frac{L}{1+L^2}
\Big)-4m\frac{1-L^2}{1+L^2}&=0~,\nn
\end{align}
}}
\end{center}
\be\label{lconstraints}\end{equation}
where here and in the remainder of this paper 
the Hodge star is taken with respect to the internal eight-dimensional space. 
Moreover, all flux components except for the $\bf{27}$ component of $F$ are 
solved for in terms of $L$ and the warp factor:
\begin{center}
\fbox{\parbox{15cm}{
%
%
\beal
f&=e^{-3\Delta}d\Big(
e^{3\Delta}\frac{1-L^2}{1+L^2}
\Big)+8m\frac{L}{1+L^2}\nn\\
\frac{1}{12}F^{\bf{1}}&=
e^{-3\Delta}L^i\partial_i
\Big(\frac{e^{3\Delta}}{1+L^2}\Big)-m\frac{3-L^2}{1+L^2}\nn\\
\frac{1}{96}F^{\bf{7}}_{rs}&=-
e^{-3\Delta}(P^{\bf{7}})^{pq}_{rs}L_p\partial_q
\Big(\frac{e^{3\Delta}}{1+L^2}\Big)\nn\\
\frac{1}{24}F^{\bf{35}}_{mn}&=
-\nabla_{(m}L_{n)}-\frac{1}{4}\Phi_{(m}{}^{ijk}(L\otimes F^{\bf{27}})^{\bf{48}}_{n)ij}L_k
+\frac{3}{7(1+L^2)^2}\Big(
L_mL_n+\frac{L^2}{6}g_{mn}
\Big)L^i\nabla_i L^2\nn\\
&-\frac{9}{7(1+L^2)}\Big\{
L_mL_n+\frac{7+8L^2}{6}g_{mn}
\Big\}L^i\nabla_i \Delta
+\frac{1}{(1+L^2)^2}L_{(m}\partial_{n)}L^2\nn\\
&+\frac{3L^2}{1+L^2}L_{(m}\partial_{n)}\Delta
+\frac{m}{14(1+L^2)}\Big\{
8(L^2-3)L_mL_n+(7+3L^2-8L^4)g_{mn}
\Big\}~,\nn
\end{align}
%
}}
\end{center}
\be\label{fluxes}\end{equation}
where the explicit decomposition of $F$ in terms of irreducible representations of 
$Spin(7)$ and the explanation of 
the definitions entering the equations above, are given in the appendix. 
As noted in the introduction, in open 
sets where neither of $\xi^{\pm}$ vanishes,  
the above equations  
 should reduce to the corresponding formul{\ae} given in \cite{ms}, to the 
extent they overlap (some of the flux components were 
not given explicitly in \cite{ms}). 
Finally, the intrinsic torsion is determined via
\begin{center}
\fbox{\parbox{8cm}{
%
%
\beal
\omega^1_m&=
\frac{m}{2}L_m+\frac{3}{4}\partial_m\Delta
+\frac{1}{168}(L_mF^{\bf{1}}-L^iF_{im}^{\bf{7}} )\nn\\
\omega^2_{mpq}&=
\frac{1}{192}(L\otimes F^{\bf{7}})^{\bf{48}}_{mpq}
+\frac{1}{4}(L\otimes F^{\bf{27}})^{\bf{48}}_{mpq}
~.\nn
\end{align}
%
}}
\end{center}
\be\label{tors}\end{equation}

\subsection{Small-flux approximation}

A special solution to the supersymmetry equations (\ref{1}-\ref{4}) the 
Bianchi identities and the equations-of-motion (\ref{eom}), is obtained when 
the warp factor and all flux vanishes 
($\Delta$, $f$, $F=0$), ${\cal M}$ is three-dimensional Minkowski space ($m=0$) 
and $X$ is a manifold of $Spin(7)$ holonomy ($\omega^{1,2}=0$). 
In this section we 
would like to perform a small perturbation around the 
special-holonomy solution; 
this amounts to 
a small-flux approximation. Note that this is an expansion 
around the point where the $G_2=Spin(7)_+\cap Spin(7)_-$ 
structure breaks down, and so 
it cannot be described by the formalism of \cite{ms}.

For each field $S$, let us 
make a perturbative expansion
\beal
S=\sum_{n=0}S^{(n)}\varepsilon^n~,
\end{align}
where $\varepsilon$ is a small parameter, so that 
the special-holonomy solution is recovered in the 
$\varepsilon\rightarrow 0$ limit. Equations (\ref{fluxes}) determine 
the flux components:
\beal
F^{\bf{1}}&=-36m^{(1)}\varepsilon+{\cal O}(\varepsilon^2)\nn\\
F_{mn}^{\bf{7}}&={\cal O}(\varepsilon^2)\nn\\
F^{\bf{35}}_{mn}&=
\Big(
-24\nabla_{(m}L^{(1)}_{n)}+12g_{mn}m^{(1)}
\Big)\varepsilon
+{\cal O}(\varepsilon^2)~,
\end{align}
where the metric and the Levi-Civita connection above are
 those of the unperturbed special-holonomy solution. 
The ${\bf{27}}$ component of the flux is of order ${\cal O}(\varepsilon)$, 
but is otherwise unrestricted by the supersymmetry equations. Moreover, 
equations (\ref{tors}) give
\beal
\omega^1_m&=
\frac{3}{4}\partial_m\Delta^{(1)}+{\cal O}(\varepsilon^2)\nn\\
\omega^2_{mpq}&={\cal O}(\varepsilon^2)~.
\end{align}
It follows from the form of the intrinsic torsion 
(see e.g. \cite{kari}) that to order 
${\cal O}(\varepsilon^2)$ the eight-manifold is conformally 
special-holonomy. Finally, to order ${\cal O}(\varepsilon^2)$, 
equations (\ref{lconstraints}) are equivalent to 
\beal
\nabla^mL^{(1)}_m&=4m^{(1)}
\nn\\
\nabla_{[m}L^{(1)}_{n]}&=0~,
\label{qo}
\end{align}
where again the Levi-Civita connection above is
that of the special-holonomy metric. 
Integrating the first line by parts in the case where
$X$ is smooth and compact, we conclude that $m^{(1)}=0$. 
By the same reasoning, it can be seen by induction that $m$ vanishes to all orders 
in $\varepsilon$. It follows that, as noted in section \ref{nsusy},  
all flux vanishes and we get back the special-holonomy solution. 


Nontrivial solutions can be obtained if $X$ is noncompact. In this case 
equations (\ref{qo}) are solved for
\beal
L^{(1)}_m&=\partial_m\phi\nn\\
\Box \phi&=4m^{(1)}~,
\label{qo1}
\end{align}
where $\phi$ is a scalar on $X$ with dimensions of length 
and the box operator is taken 
with respect to the unperturbed special-holonomy metric. The Bianchi 
identities and  the equations-of-motion 
impose the conditions that $\Delta^{(1)}$ should 
be harmonic and that $F^{\bf{27}}$ should be closed. 
Neglecting corrections 
of order ${\cal O}(\varepsilon^2)$, it can be shown that 
there are no further conditions
\footnote{
To arrive at this 
result the only nontrivial step is to prove 
that $I_{m_1\dots m_5}:=\Phi_{[m_1\dots m_3}{}^pR_{m_4m_5],p}{}^qL_q$ vanishes, where 
the Riemann tensor is with respect to the special-holonomy connection. 
This can be seen as follows: for fixed $m,n$, $R_{mn,pq}$ can be viewed as 
an antisymmetric matrix with indices $p,q$; i.e. 
it transforms in the $\bf{21}+\bf{7}$ 
of $spin(7)$. However, since 
the spinor $\eta$ is parallel with respect to the connection, 
it follows that $R_{mn,pq}\ga^{pq}\eta=0$ and hence 
the $\bf{21}$ component is projected out  
while the $\bf{7}$ component is set to zero (cf. equation (\ref{21})). I.e.  
for fixed $m,n$ (or for fixed $p,q$, thanks to 
the symmetry properties of the Riemann tensor) $R_{mn,pq}$ 
transforms in the $\bf{21}$ of $spin(7)$. Furthermore, 
as follows from the symmetry of the free indices and the previous discussion, 
$R_{mn,p}{}^qL_q$ transforms in the 
$\bf{21}\otimes \bf{8}=\bf{8}\oplus\bf{48}\oplus\bf{112}$. On the other hand, 
$I_{m_1\dots m_5}$ is a five-form and hence it transforms in the 
$\bf{8}\oplus\bf{48}$ of $spin(7)$; therefore the $\bf{112}$ representation 
is projected out. The remaining $\bf{8}\oplus\bf{48}$ representations are 
generated by $R_{[mnp]}{}^qL_q$, which 
vanishes by virtue of the symmetries of the Riemann 
tensor, and $g_{mn}R_{p}{}^qL_q$, which vanishes 
by virtue of the fact that 
every manifold of special holonomy is Ricci-flat. It follows that 
$I_{m_1\dots m_5}$ vanishes.}.

As an explicit example, let us consider 
a small perturbation around the noncompact $Spin(7)$-holonomy metric 
of reference \cite{sala, gibb}:
\beal
ds^2=\Big(1-\Big(\frac{l}{r}\Big)^{10/3} \Big)^{-1}dr^2
+\frac{9}{100}r^2\Big(1-\Big(\frac{l}{r}\Big)^{10/3} \Big)(\sigma_i-A^i)^2
+\frac{9}{20}r^2d\Omega_4^2
~,
\end{align}
where $l\leq r<\infty$, $d\Omega_4^2$ is the metric of the unit four-sphere $S^4$, 
$\{\sigma_i; ~ i=1,2,3\}$ are left-invariant $SU(2)$ one-forms and $A^i$ 
is the connection of a Yang-Mills instanton on $S^4$. 
Moreover, let us assume that $\phi=\phi(r)$. 
For $r>>l$, the solution to equation (\ref{qo1}) behaves as 
$$
\phi\sim Q_1+Q_2\frac{1}{\delta^2}+{\cal O}(\delta)~,
$$
where $\delta:=l/r$ and $Q_{1,2}$ are constants ($Q_2$ depends on $m$). 
On the other hand, when $r$ approaches $l$ we have 
$$
\phi\sim Q'_1+Q'_2\Big(
\frac{1}{\delta}+\frac{8}{3}{\rm log}(\delta)
\Big)+{\cal O}(\delta)~,
$$
where $Q'_{1,2}$ are constants ($Q'_2$ depends on $m$) and $\delta:=r/l-1$.  
In conclusion: for $m\neq 0$ (i.e. for ${\cal M}$ AdS), 
$L^2\sim |\partial\phi|^2$ blows up near $l$, $\infty$, which 
is contrary to our assumption that $L$ is perturbatively small.  
Hence, the solution can only 
be trusted for intermediate distances $1/\varepsilon>>r/l>> \varepsilon$. 
Note however that for $m=0$ (i.e. for ${\cal M}$ Minkowski) 
$Q_2$ vanishes and the solution to (\ref{qo1}) 
is regular for large distances $r/l>> \varepsilon$.

\section{Generalized G-structures}
\label{generalized}

In this section, after some preliminaries, we give the definition of 
generalized $Spin(7)$ structures in nine dimensions and 
explain how they arise naturally in the context of supersymmetric 
M-theory compactifications on eight-manifolds.  Generalized $G$-structures 
were first introduced in \cite{hitchin}. 
Generalized $Spin(7)$ structures in eight dimensions were first examined 
by F.~Witt in \cite{1}.
 
Consider the direct sum of the tangent and cotangent bundle $T\oplus T^{*}$ of 
a $d$-dimensional manifold $X$. 
There is a natural action of $T\oplus T^{*}$ on forms, whereby every vector acts by 
contraction and every form by exterior multiplication. Explicitly: if 
$V$ is a vector on $X$ and $U$, $\Omega$ are forms,  
we define
\beal
(V+U)\cdot \Omega= \iota_V\Omega+U\wedge\Omega~.
\end{align}
As this action squares to the identity, there is an associated Clifford algebra 
${\rm Cliff}(T\oplus T^{*})$ and an induced isomorphism 
\beal
{\rm Cliff}(T\oplus T^{*})\thickapprox {\rm End}(\Lambda^{*})~.
\label{iso}
\end{align}
A basis of the Clifford algebra on $T\oplus T^{*}$ 
\beal
\{\ga^m,\ga^n\}=0;~~~~~\{\ga_m,\ga_n\}=0;~~~~~\{\ga^m,\ga_n\}=\delta^m_n
\end{align}
is given explicitly by $\ga^m:=dx^m\wedge, ~\ga_n:=\iota_n$. 
It follows from the isomorphism (\ref{iso}) above that (sums of) forms on $X$ 
can be identified with spinors of $T\oplus T^{*}$. Moreover, the latter 
can be thought of as bispinors on $X$. We thus obtain
\beal
{\rm forms}\longleftrightarrow {\rm spinors~on}~ T\oplus T^{*} 
\longleftrightarrow {\rm bispinors~on}~ X~.
\end{align}
The identification of sums of forms with bispinors is, of course, 
explicitly realized by Fierzing.

\subsection{Generalized Spin(7) structures}
\label{generalizeds}

Coming back to our eight-dimensional case, 
we define the following bispinors
\beal
\Phi^{\pm}&:=\xi^{\pm}\otimes\xi^{\pm}=\frac{1}{8}P_{\pm}\Big\{ 
\Phi^{\pm}+\frac{1}{2\cdot 4!}\Phi^{\pm}_{m_1\dots m_4}\ga^{m_1\dots m_4}
+\frac{1}{8!}\Phi^{\pm}_{m_1\dots m_8}\ga^{m_1\dots m_8}
\Big\}~,
\end{align}
where $P_{\pm}:=\frac{1}{2}(1\pm\ga_9)$ is the chirality projector and 
$\Phi_{m_1\dots m_p}^{\pm}:=\xi^{\pm}\ga_{m_1\dots m_p}\xi^{\pm}$. 
By a slight abuse of notation, 
we use the same letter to denote both the bispinor and the associated forms. It should 
be clear from the context which one is meant in each case.
It will prove more convenient to work with the 
linear combinations: $\Psi^{\pm}:=\frac{1}{2}(\Phi^+\pm\Phi^-)$. 
Moreover we define
\beal
\wps&:=\xi^{+}\otimes\xi^{-}=\frac{1}{8}P_+\sum_{p={\rm odd}}
\frac{1}{p!}\wps_{m_1\dots m_p}\ga^{m_1\dots m_p}
~, 
\end{align}
where $\wps_{m_1\dots m_p}:=\xi^{+}\ga_{m_1\dots m_p}\xi^{-}$. In 
the following we will find it useful to define the 
combinations (cf. also equations
(\ref{p1},\ref{p2}) of appendix \ref{derof}): $\wps^{\pm}:=\frac{1}{2}(\wps\pm\star\wps)$.

The Majorana spinors 
(real pinors) $\epsilon^{\pm}:=\frac{1}{\sqrt{2}}(\xi^+\pm \xi^-)$ 
are nowhere-vanishing on $X$, since $2|\epsilon|^2=|\xi^+|^2+|\xi^-|^2=1$. 
Hence any bipinors constructed out of the $\epsilon$'s 
are also nowhere-vanishing. Setting
\beal
\rho^{\pm}&:=\epsilon^{\pm}\otimes\epsilon^{\pm}=
\psp\pm\wps^-\nn\\
\widehat{\rho}^{\pm}&:=\epsilon^{\pm}\otimes\epsilon^{\mp}
=\psm\mp\wps^+
~,
\end{align}
we note that $\Psi^{\pm}$, 
$\wps^{\pm}$ can be expressed as linear combinations of $\rho, \widehat{\rho}$. 
As discussed in detail in section \ref{spinstructure}, 
these bipinors on $X$ lift to 
bispinors on $Y=X\times S^1$. 
We shall call the pair ($\rho,~\widehat{\rho}$) a 
{\it generalized} $Spin(7)$ {\it structure on} $Y$. Note that 
   ($\rho,~\widehat{\rho}$) induce a reduction of the structure 
group $Spin(9,9)$ of $TY\oplus T^*Y$ to $Spin(7)\times Spin(7)$. This follows 
from the fact that $\epsilon^{\pm}$ are nowhere-vanishing and therefore each of them 
induces a reduction of the structure group of $Y$ to $Spin(7)$, as explained in 
\ref{spinstructure}.

As we show in appendix \ref{derof}, ${\cal N}=1$ supersymmetry 
implies that 
the generalized $Spin(7)$ structure, or equivalently the 
bispinors $\Psi^{\pm}, \wps^{\pm}$,  satisfy the following 
differential equations
\begin{center}
\fbox{\parbox{10.7cm}{
%
\beal
0&=d\psp+F\wedge\wps^-\nn\\
0&=e^{-3\Delta}d(e^{3\Delta}\psm)+\star F\wedge\wps^-
-f\wedge\psp+4m\wps^-\nn\\
0&=e^{-3\Delta/2}d(e^{3\Delta/2}\wps^-)\nn\\
0&=e^{-3\Delta/2}d(e^{3\Delta/2}\wps^+)+2(\star F\wedge \psp-F\wedge\psm)
+8m\psp
~.\nn
\end{align}
%
}}
\end{center}
\be\label{esp}\end{equation}
In the terminology of \cite{jescb, jesca}, the equations (\ref{esp}) 
above are the `form picture' of the 
${\cal N}=1$ supersymmetry equations given in the 
`spinor picture' in (\ref{1}-\ref{4}). 
Note that, as is easy to show, the integrability of (\ref{esp}) follows from the 
equations-of-motion and the Bianchi identities.

\subsection{Reduction to seven dimensions}

In the case where $X$ is of the form $Z\times {S^1}$ 
and assuming no fields depend on the coordinate of $S^1$, we can 
perform a  reduction to seven dimensions --upon which 
$\xi^{\pm}$ and $\epsilon^{\pm}$ transform in the $\bf{8}$ of $Spin(7)$. 
Since as we noted above $\epsilon^{\pm}$ are nowhere-vanishing, 
each of them gives rise to a 
$G_{2\pm}\subset Spin(7)$ 
structure on $Z$. Indeed, $G_2$ is the isotropy group of a 
fundamental spinor inside $Spin(7)$. Alternatively, this can be seen 
by noting that under $G_2\subset Spin(7)$ the fundamental 
spinor representation decomposes as $\bf{8}\rightarrow \bf{7}+\bf{1}$, i.e. 
there is a singlet in the decomposition. 
If $\epsilon^{\pm}$ are nowhere parallel 
(equivalently: if $\xi^{\pm}$ are nowhere-vanishing), there is a further 
reduction of the structure group of $Z$ to the common subgroup 
$G_{2+}\cap G_{2-}=SU(3)$. In the generic case there are points in $Z$ where 
$\epsilon^{\pm}$ become parallel. At these points 
the $SU(3)$ structure is enhanced to $G_{2+}\cap G_{2-}=G_2$. This 
situation is best described in 
the language of 
generalized $G_2$ structures in seven dimensions \cite{jescb, jesca}.

\section{Conclusions}

We have presented a formalism for 
supersymmetric M-theory compactifications on 
eight-manifolds $X$, based on the group $Spin(7)$. 
This is the most suitable language in which 
to describe compactifications on eight-manifolds of 
$Spin(7)$ structure, and/or small-flux perturbations 
around compactifications on manifolds of $Spin(7)$ 
holonomy. Although supersymmetry does not, in general, 
imply the reduction of the structure group of $X$ itself, 
our analysis leads naturally to the emergence of a nine-dimensional 
manifold $Y=X\times S^1$ whose structure group is reduced to 
$Spin(7)$. This is reminiscent of the connection between M- and F-theory: 
in the case where $X$ admits an elliptic fibration, M-theory on $X$ is 
equivalent to F-theory on $X\times S^1$ \cite{vafa}. 
It would be very interesting to explore this similarity further.

In eight dimensions  there exists a 
Hitchin functional involving a certain three-form and 
its Hodge-dual five-form \cite{hitb}. This does 
not seem to be related to the case considered here: we generally 
do not have any nowhere-vanishing three-form on $X$. 
It would be interesting to explore whether 
such a functional can be constructed using the four-forms 
$\Phi^{\pm}$ of section \ref{generalizeds} and what should 
be the generalization of stability in this case. 
A related point is that,  
as we have already noted in the introduction, in ten dimensions 
the NS and RR fields play quite different roles 
with respect to the Hitchin functional construction \cite{jesca};
 this distinction disappears upon lifting to M-theory. 
It would also be desirable to know whether 
equations (\ref{esp}) can be interpreted 
as some sort of 
integrability  condition for the generalized $Spin(7)$ structures.

In type II theories the generalized picture provides a natural framework 
for T-duality in topological models \cite{kapu, jescc,  zuch}.  It would 
be interesting to explore this issue in the context corresponding 
to the setup of this paper, i.e. in the context of an M- or F-theory 
topological $\sigma$-model with target space the eight-manifold $X$.  
We expect T-duality to act as a sign flip: $\xi^-\rightarrow -\xi^-$. 
This amounts to an exchange $\epsilon^+\leftrightarrow\epsilon^-$ 
(cf. section \ref{generalizeds}), in 
analogy to the situation in seven and six dimensions.

At a more mundane level, 
the present paper opens up the possibility for 
supersymmetric solutions with all fluxes 
turned on and with an internal 
manifold in any of the four classes of 
$Spin(7)$ manifolds. Any explicit examples of such 
manifolds are, of course, desirable and 
could provide us with  interesting physics; 
the physics of M theory on eight manifolds is already 
very rich, even in the case where the internal manifold 
is special-holonomy. Even in the absence 
of an explicit metric, the characterization of the 
most general ${\cal N}=1$ backgrounds given in this paper 
should suffice for a Kaluza-Klein reduction and the derivation 
of the resulting low-energy supergravity in three dimensions. 
It will be interesting to pursue this point further.

\section{Acknowledgment}

I would like to thank Claus Jeschek for valuable discussions 
and Dario Martelli 
for email correspondence. I am also grateful to Frederik Witt for 
useful discussions and explanations, and to the referee of JHEP for 
useful suggestions.

\appendix

\section{Gamma-matrix identities in 8d}

The gamma matrices in eight dimensions have the following properties

Symmetry:
\beal
(C\ga_{(n)})^{Tr}=(-)^{\frac{1}{2}n(n-1)}C\ga_{(n)}~,
\end{align}
where $C$ is the charge-conjugation matrix.

Hodge-duality:
\beal
\star\ga_{(n)}=(-)^{\frac{1}{2}n(n+1)}\ga_{(8-n)}\ga_9~,
\end{align}
where $\ga_9$ is the chirality matrix.

\section{Identities relating to the $Spin(7)$ structure}

In this section we give a number of identities 
which we have used repeatedly in this paper. 
These can be proved either by Fierzing or by 
fixing a special basis for the spinor $\eta$, as in e.g. \cite{gp}.  
Given a positive-chirality Majorana spinor $\eta$ of unit norm, we can define 
a self-dual four-form as in (\ref{s7}), which can be seen to satisfy the 
following identities
\beal
\Phi^{ijkl}\Phi_{ijkm}&=42\delta^l_m\nn\\
\Phi^{ijkl}\Phi_{ijpq}&=12\delta^{kl}_{pq}-4\Phi^{kl}{}_{pq}\nn\\
\Phi^{iklm}\Phi_{ipqr}&=6\delta^{klm}_{pqr}-9\Phi^{[kl}{}_{[pq}\delta^{m]}_{r]} ~.
\end{align}
Moreover, we have
\beal
\ga_{ij}\eta&=-\frac{1}{6}\Phi_{ij}{}^{kl}\ga_{kl}\eta\nn\\
\ga_{ijk}\eta&=-\Phi_{ijk}{}^l\ga_l\eta\nn\\
\ga_{ijkl}\eta&=\Phi_{[ijk}{}^m\ga_{l]m}\eta+\Phi_{ijkl}\eta\nn\\
\ga_{ijklm}\eta&=5\Phi_{[ijkl}\ga_{m]}\eta~.
\label{b2}
\end{align}
We define the following projectors, 
acting on a second-rank tensor, 
onto the 
$\bf{7}$, $\bf{35}$ of $Spin(7)$:
\beal
(P^{\bf{7}})_{mn}^{pq}
:=\frac{1}{4}\Big(\delta_{[m}^p\delta^q_{n]}-\frac{1}{2}\Phi_{mn}{}^{pq} \Big) 
\label{p7}\\
(P^{\bf{21}})_{mn}^{pq}
:=\frac{3}{4}\Big(\delta_{[m}^p\delta^q_{n]}+\frac{1}{6}\Phi_{mn}{}^{pq} \Big) \\
(P^{\bf{35}})_{mn}^{pq}:=  \delta_{(m}^p\delta^q_{n)}-\frac{1}{8}g_{mn}g^{pq}
\label{p35}~.
\end{align}
A useful identity is
\beal
(P^{\bf{21}})_{rs}^{pq}\ga_{pq}\eta=0~.
\label{21}
\end{align}

\section{$Spin(7)$ tensor decomposition}

Let us decompose $F_{mnpq}$ into irreducible representations
\beal
F_{mnpq}=F_{mnpq}^{\bf{1}}+F^{\bf{7}}_{mnpq}+F^{\bf{27}}_{mnpq}
+F^{\bf{35}}_{mnpq}~.
\end{align}
Expanding
\beal
F^{\bf{1}}_{mnpq}&=\frac{1}{42}\Phi_{mnpq}F^{\bf{1}}\nn\\
F^{\bf{7}}_{mnpq}&=\frac{1}{24}\Phi_{[mnp}{}^iF^{\bf{7}}_{q]i}\nn\\
F^{\bf{35}}_{mnpq}&=\frac{1}{6}\Phi_{[mnp}{}^iF^{\bf{35}}_{q]i}~,
\label{kiri}
\end{align}
where $F^{\bf{7}}_{mn}$ is antisymmetric in $m, n $ whereas $F^{\bf{35}}_{mn}$
is symmetric and traceless, we obtain
\beal
F_{ijkp}\Phi^{ijk}{}_q=g_{pq}F^{\bf{1}}+F^{\bf{7}}_{pq}+F^{\bf{35}}_{pq}~.
\end{align}
In the above we have noted that 
\beal
F^{\bf{27}}_{ijkp}\Phi^{ijk}{}_q=0~.
\end{align}
This can be seen immediately as follows: the left-hand side transforms in 
the $\bf{8}\otimes\bf{8}$ of $Spin(7)$, however  
$\bf{8}\otimes\bf{8}= \bf{1}\oplus\bf{7}\oplus\bf{21}\oplus\bf{35}$ and there is no 
$\bf{27}$ in the decomposition. 
Note that it follows from decomposition (\ref{kiri}) that $F^{\bf{1}}$, $F^{\bf{7}}$
 are self-dual while $F^{\bf{35}}$ is anti self-dual.

In deriving (\ref{1.2}) below, we shall need the following decompositions. 
\beal
L_mF^{\bf{7}}_{pq}&= (P^{\bf{7}})_{pq}^{ij}\Big\{  
(L\otimes F^{\bf{7}})_{mij}^{\bf{48}}+g_{mi}(L\otimes F^{\bf{7}})_j^{\bf{8}}
\Big\} \\
L_mF^{\bf{35}}_{pq}&=(P^{\bf{35}})_{pq}^{ij}\Big\{  
\Phi_{mi}{}^{kl}(L\otimes F^{\bf{35}})_{jkl}^{\bf{48}}+g_{mi}(L\otimes F^{\bf{35}})_j^{\bf{8}}
\Big\}+\dots~,
\end{align}
where the ellipses stand for the irreducible representations which drop out of (\ref{1.2}). 
These expansions can be inverted to give
\beal
(L\otimes F^{\bf{7}})_m^{\bf{8}}&=\frac{8}{7}L^iF^{\bf{7}}_{im}\\
(L\otimes F^{\bf{7}})_{mpq}^{\bf{48}}&=6\Big(
L_{[m}F^{\bf{7}}_{pq]}+\frac{1}{7}\Phi_{mpq}{}^jL^iF^{\bf{7}}_{ij}
\Big)
\label{b8}
\end{align}
and
\beal
(L\otimes F^{\bf{35}})_m^{\bf{8}}&=\frac{8}{35}L^iF^{\bf{35}}_{im}\\
(L\otimes F^{\bf{35}})_{mpq}^{\bf{48}}&=\frac{3}{20}\Big(
L_{i}F^{\bf{35}}_{j[m}\Phi_{pq]}{}^{ij}-\frac{1}{7}\Phi_{mpq}{}^jL^iF^{\bf{35}}_{ij}
\Big)
~.
\label{b10}
\end{align}
The reader can verify that the right-hand sides of (\ref{b8}, \ref{b10}) transform 
in the $\bf{48}$ of $Spin(7)$, as they should. Moreover, note that 
\beal
(L\otimes F^{\bf{27}})_{mpq}^{\bf{48}}:=L^iF^{\bf{27}}_{impq}
\end{align}
also transforms in the $\bf{48}$. Indeed the right-hand 
side is a three-form and therefore transforms in the 
$\bf{8}^{3\otimes_a}=\bf{8}\oplus\bf{48}$. On the other hand, the 
right-hand side is the product of $L$, $F^{\bf{27}}$, and therefore transforms in the 
$\bf{8}\otimes\bf{27}=\bf{48}\oplus\bf{168}$. It follows that the right-hand side 
is in the $\bf{48}$ of $Spin(7)$.

\section{${\cal N}=1$ supersymmetry}
\label{susyd}

In this appendix we  give the details 
of the derivation of equations (\ref{x}, \ref{lconstraints}-\ref{tors}). 
Taking decomposition (\ref{kiri}) into account,  
the supersymmetry  
transformations (\ref{1}-\ref{4}) can be seen to be 
 equivalent to the following conditions. 

Equation (\ref{1}):
\beal
0&=\partial_mR-(m+\frac{1}{24}F^{\bf{1}})L_m+\frac{1}{24}L^{i}(F^{\bf{7}}_{im}
-F^{\bf{35}}_{im})
\label{1.1}\\
0&=(P^{\bf{7}})_{rs}^{pq}\Big\{g_{mp}(\omega^1_q-\frac{1}{4}f_q+mL_q)
-\omega^2_{mpq}+\frac{1}{24}(L_p\Phi_q{}^{ijk}F_{mijk}+6L^iF_{impq})\Big\}
~,
\label{1.2}
\end{align}
where we have parameterized $\xi^+=e^{R}\eta$, $\xi^-=e^RL_m\gamma^m\eta$. 
Equation (\ref{2}):
\beal
0&=\nabla_mL_n+\partial_mRL_n
+\Phi_n{}^{ijk}\omega^2_{mij}L_k-2\omega^2_{mni}L^i
+g_{mn}(L^i\omega^1_i-\frac{1}{4}L^if_i+\frac{1}{24}F^{\bf{1}}+m)\nn\\
&-L_m(\omega^1_n-\frac{1}{4}f_n)
-\Phi_{mn}{}^{ij}L_i(\omega^1_j+\frac{1}{4}f_j)
+\frac{1}{24}(F^{\bf{7}}_{mn}+F^{\bf{35}}_{mn})
~.
\label{2.1}
\end{align}
Equation (\ref{3}):
\beal
0&=mL_m+\frac{1}{2}(\partial_m\Delta-\frac{1}{3}f_m)
+\frac{1}{36}L^iF^{\bf{35}}_{im}
\label{3.1}
~.
\end{align}
Equation (\ref{4}):
\beal
0&=m-\frac{1}{2}L^i(\partial_i\Delta+\frac{1}{3}f_i)
+\frac{1}{36}F^{\bf{1}}
\label{4.1}\\
0&=
(P^{\bf{7}})_{rs}^{pq}
\Big\{L_p\partial_q\Delta 
+\frac{1}{3}L_pf_q
\Big\}+\frac{1}{144}F^{\bf{7}}_{rs}
~.
\label{4.2}
\end{align}

Before proceding to the derivation of 
(\ref{1.1}-\ref{4.2}), let us mention that 
equation (\ref{x}) is derived as follows: 
multiplying (\ref{2.1}) by $L^n$ 
and using (\ref{1.1}), we arrive at
\beal
\partial_mR=-\frac{L^n}{1+L^2}\nabla_mL_n~.
\end{align}
Taking into account that $\xi^{\pm}$ can be rescaled by a 
real constant without loss
 of generality, it follows that 
\beal
e^R=\frac{1}{\sqrt{1+L^2}}~,
\label{r}
\end{align}
which is equivalent to equation (\ref{x}).

In deriving equations (\ref{1.1}, \ref{1.2}, \ref{4.1}, \ref{4.2}) we have noted that the equation
\beal
A_m\eta+B_{m,pq}\ga^{pq}\eta=0~,
\end{align}
where $(P^{\bf{7}})_{pq}^{rs}B_{m,rs}=B_{m,pq}$, 
is equivalent to $A_m,\,\, B_{m,pq}=0$. 
This can be seen by multiplying on the left by $\eta$ 
and $\eta\ga^{rs}$. Similarly, in deriving (\ref{2.1}, \ref{3.1}) we have noted that 
the equation 
\beal
A_{m,n}\ga^{n}\eta=0~
\end{align}
is equivalent to $A_{m,n}=0$, as can be seen by multiplying on the left 
by $\eta\ga_p$.

Equation (\ref{1.2}) can be used to solve for the intrinsic torsion
as in (\ref{tors}), by taking into account the following identities
\beal
0&=(P^{\bf{7}})_{rs}^{pq}\Big\{
\Phi_{pq}{}^{ab}+6\delta_{[p}^a\delta_{q]}^b
\Big\}\\
0&=(P^{\bf{7}})_{rs}^{pq}\Big\{
L_pF_{mq}^{\bf{7}}-\frac{1}{4}(L\otimes F^{\bf{7}})^{\bf{48}}_{mpq}+\frac{5}{7}g_{mp}L^iF^{\bf{7}}_{iq}
\Big\}\\
0&=(P^{\bf{7}})_{rs}^{pq}\Big\{
L^i\Phi_{imp}{}^jF_{qj}^{\bf{7}}+\frac{1}{4}(L\otimes F^{\bf{7}})^{\bf{48}}_{mpq}
-\frac{12}{7}g_{mp}L^iF^{\bf{7}}_{iq}
\Big\}\\
0&=(P^{\bf{7}})_{rs}^{pq}\Big\{
L_pF_{mq}^{\bf{35}}+5(L\otimes F^{\bf{35}})^{\bf{48}}_{mpq}-\frac{1}{7}g_{mp}L^iF^{\bf{35}}_{iq}
\Big\}\\
0&=(P^{\bf{7}})_{rs}^{pq}\Big\{
L^i\Phi_{imp}{}^jF_{qj}^{\bf{35}}+5(L\otimes F^{\bf{35}})^{\bf{48}}_{mpq}
+\frac{6}{7}g_{mp}L^iF^{\bf{35}}_{iq}
\Big\}\\
0&=(P^{\bf{7}})_{rs}^{pq}\Big\{
L_p\Phi_q{}^{ijk}F_{mijk}+L_iF_{jkmp}\Phi_q{}^{ijk}+4L^iF_{impq}
\Big\} ~.
\end{align}
In order to solve for the $F^{\bf{35}}$ component of $F$, we first define
\beal
f^{\bf{35}}_{mn}:=F^{\bf{35}}_{mn}
&+2\Big\{L^iF^{\bf{35}}_{i(m}L_{n)}-\frac{1}{8}g_{mn}L^iL^jF^{\bf{35}}_{ij}\Big\}\nn\\
&+\frac{6}{7}L^iL^jF^{\bf{35}}_{ij}\Big( 
L_mL_n-\frac{1}{8}g_{mn}L^2
\Big)~,
\label{yu}
\end{align}
which is the combination that appears in the symmetric, traceless part of (\ref{2.1}).  
Note that $f^{\bf{35}}$ indeed 
transforms in the $\bf{35}$ of $Spin(7)$. Equation (\ref{2.1}) can then be used to solve for 
$f^{\bf{35}}$:
\beal
\frac{1}{24}&f^{\bf{35}}_{mn}+\Big(\nabla_{(m}L_{n)}-\frac{1}{8}g_{mn}\nabla L   \Big)
+\frac{1}{4}\Phi_{(m}{}^{ijk}(L\otimes F^{\bf{27}})^{\bf{48}}_{n)ij}L_k\nn\\
&+\frac{5}{7}\Big(L_mL_n-\frac{1}{8}g_{mn}L^2\Big)
\Big\{
m(1+\frac{9}{5}L^2)+\frac{9}{5}L^i\partial_i\Delta
\Big\}
=0~,
\end{align}
which can be solved for $F^{\bf{35}}$ by inverting  
(\ref{yu}):
\beal
F^{\bf{35}}_{mn}&=f^{\bf{35}}_{mn}-\frac{2L^if^{\bf{35}}_{i(m}L_{n)}}{1+L^2}\nn\\
&+\frac{L^iL^jf^{\bf{35}}_{ij}}{(1+L^2)(1+\frac{3}{4}L^2)}
\Big\{
\frac{9}{14}L_mL_n+\frac{1}{4}g_{mn}(1+\frac{3}{7}L^2)
\Big\}~.
\end{align}
Note that $F^{\bf{35}}_{mn}$ in (\ref{fluxes}) is traceless by virtue of
\beal
\Big\{
g^{mn}-2\frac{L^mL^n}{(1+L^2)}\Big\}\nabla_mL_n-
4m(1-L^2)+12L^i\partial_i\Delta=0
~,
\label{poui}
\end{align}
which is obtained by tracing equation (\ref{2.1}). 
A straightforward manipulation then leads to the second line of equation (\ref{lconstraints}).

The $\bf{7}$ and $\bf{21}$ parts of (\ref{2.1}) are treated similarly. 
Taking the identities
\beal
0&=(P^{\bf{7}})_{rs}^{pq}\Big\{
\Phi_{p}{}^{ijk}\omega^2_{qij}L_k+4\omega^2_{pqi}L^i
\Big\}\nn\\
0&=(P^{\bf{21}})_{rs}^{pq}\Big\{
\Phi_{p}{}^{ijk}\omega^2_{qij}L_k
\Big\}
\end{align}
into account, 
we arrive at
\beal
(\nabla_{[r}L_{s]})^{\bf{7}}&=(P^{\bf{7}})^{mn}_{rs}\Big\{
L^i\nabla_iL_mL_n-\frac{1}{2}L_m\partial_nL^2
+3(1+L^2)L_m\partial_n\Delta
\Big\}\\
(\nabla_{[r}L_{s]})^{\bf{21}}&=(P^{\bf{21}})^{mn}_{rs}\Big\{
L^i\nabla_iL_mL_n-\frac{1}{2}L_m\partial_nL^2
+3(1+L^2)L_m\partial_n\Delta
\Big\}
~.
\end{align}
It is then straightforward to show that the equations above are equivalent to the first 
line of (\ref{lconstraints}). 
Taking all the above into account, the expressions for the remaining flux components 
$f$, $F^{\bf{1}}$, $F^{\bf{7}}$ are obtained by straightforward 
manipulations of equations (\ref{3.1}, \ref{4.1}, \ref{4.2}), respectively.

\section{Spinor vs four-form}
\label{equivalence}

In this section we prove the equivalence of the two equations in (\ref{a}). 
One needs to show that the existence of 
the self-dual four-form $\Phi$ is equivalent 
to the existence of the chiral spinor $\eta$ (see e.g. \cite{joyce}). 
The two equations in (\ref{a}) are then essentially 
equivalent to the statement that $\Phi$, $\eta$ are acted upon by the 
Levi-Civita, the (associated) spin connection respectively, and that they  
are both $Spin(7)$ singlets. More explicitly,

$(\Longrightarrow)$: Let us expand
\beal
\nabla_m\eta=A_m\eta+B_{m,pq}\ga^{pq}\eta~,
\label{kra}
\end{align}
where without lost of generality, 
as can be seen from the first line of 
equation (\ref{b2}),  $B_{m,pq}$ can be taken to satisfy
\beal
B_{m,pq}=(P^{\bf{7}})_{pq}^{rs}B_{m,rs}~.
\label{q}
\end{align}
From (\ref{q}) we can see immediately that
\beal
4B_{[m,pq]}&=B_{i,j[m}\Phi^{ij}{}_{pq]}-\Phi_{mpq}{}^jB^i{}_{,ij}\\
B_{i,jk}\Phi^{ijk}{}_m&=-6B^i{}_{,im}~,
\label{ra}
\end{align}
from which it follows that
\beal
B_{m,pq}\ga^{pq}\eta=6B_{[m,pq]}\ga^{pq}\eta+4B^p{}_{,pq}\ga^q{}_m\eta~.
\end{align}
Moreover, the fact that $\eta$ 
has unit norm implies $A_m=0$. 
On the other hand, it follows from the first line in (\ref{a}) and (\ref{kra}) that 
\beal
\omega^1_m&=-\frac{8}{7}B^i{}_{,im} \\
\omega^2_{mpq}&=6(B_{[m,pq]}+\frac{1}{7}\Phi_{mpq}{}^jB^i{}_{,ij}  )  ~.
\label{i}
\end{align}
Taking (\ref{ra}) into account, we 
can see that the right-hand side of (\ref{i}) transforms in the $\bf{48}$ 
of $Spin(7)$, as of course it should. 
Collecting all the above, the second line of equation (\ref{a}) follows.

$(\Longleftarrow)$: 
It can be shown that the four-form $\Phi$ satisfies the following useful identity \cite{gp}
\beal
\frac{1}{24}\varepsilon_{mnpq}{}^{ijkl}=
\frac{1}{168}\Phi_{mnpq}\Phi^{ijkl}
+\frac{3}{28}\Phi_{[mn}{}^{[ij}\Phi^{kl]}{}_{pq]}
+\frac{2}{21}\Phi^{[i}{}_{[mnp}\Phi_{q]}{}^{jkl]}~.
\label{eps}
\end{align}
Moreover, any $S_{mnp}$ in the $\bf{48}$ of $Spin(7)$ satisfies
\beal
S_{mnp}&=\frac{3}{2}\Phi^{ij}{}_{[mn}S_{p]ij}\nn\\
\Phi_m{}^{ijk}S_{ijk}&=0~.
\label{48}
\end{align}
Using (\ref{eps},\ref{48}), we can see that
\beal
\varepsilon_{mnpqr}{}^{ijk}\omega^2_{ijk}=60\Phi_{[mnp}{}^i\omega^2_{qr]i}~.
\label{cookie}
\end{align}
Contracting the second line of (\ref{a}) with $\eta\ga_{pqrs}$ and using (\ref{cookie}, \ref{s7}), 
the first line of equation (\ref{a}) follows.

\section{Generalized $Spin(7)$ structures}
\label{derof}

In this appendix we include the details of the derivation 
of equation (\ref{esp}). 
Multiplying (\ref{3}), (\ref{4}) on the left by $\xi^+\ga_m$, $\xi^-\ga_m$ 
respectively and subtracting, we obtain
\beal
F_{mijk}\widehat{\Psi}^{ijk}=36\Psi^-\partial_m\Delta-12\Psi^+f_m
+72m\wps_m~.
\label{e1}
\end{align}
Multiplying (\ref{1}), (\ref{2}) on the left by $\xi^+$, $\xi^-$ respectively 
and adding/subtracting, taking (\ref{e1}) into account, we obtain
\beal
0&=\partial_m\Psi^+\nn\\
0&=e^{-3\Delta}\partial_m(e^{3\Delta}\Psi^-)-\Psi^+f_m
+4m\wps_m
~.
\label{bisp1}
\end{align}
Multiplying (\ref{3}), (\ref{4}) on the left by $\xi^+\ga_{mpqrs}$, $\xi^-\ga_{mpqrs}$ 
respectively and adding/subtracting, we obtain
\beal
\wps_{[mpq}{}^{ij}F_{rs]ij}&=
-6\psp_{[mpqr}\partial_{s]}\Delta+2
\psm_{[mpqr}f_{s]}
+\wps_{[m}F_{pqrs]}\nn\\
\wps_{[mp}{}^{i}F_{qrs]i}&=
-3\psm_{[mpqr}\partial_{s]}\Delta+\psp_{[mpqr}f_{s]}
+\frac{1}{12}\wps_{[mpqr}{}^{ijk}F_{s]ijk}-\frac{6}{5}m\wps_{mpqrs}
~.
\label{e3}
\end{align}
Multiplying (\ref{1}), (\ref{2}) on the left by $\xi^+\ga_{pqrs}$, $\xi^-\ga_{pqrs}$ 
respectively and adding/subtracting, taking (\ref{e3}) into account as well as the identity
\beal
\wps_{[mpqr}{}^{ijk}F_{s]ijk}=-6\wps_{[m}(\star F)_{pqrs]}
~, 
\end{align}
we obtain
\beal
0&=e^{-6\Delta}\partial_{[m}(e^{6\Delta}\psp_{pqrs]})+F_{[mpqr}\wps_{s]}\nn\\
0&=e^{-9\Delta}\partial_{[m}(e^{9\Delta}\psm_{pqrs]})+(\star F)_{[mpqr}\wps_{s]}
-\psp_{[mpqr}f_{s]}
+\frac{8}{5}m\wps_{mpqrs}
~.
\label{bisp2}
\end{align}
Rescaling the metric $g_{mn}\rightarrow g'_{mn}:=e^{-3\Delta}g_{mn}$ has the effect that the 
gamma matrices also get rescaled as: $\ga^{m}\rightarrow e^{3\Delta/2}\ga^{m}$.  
Passing to the bispinor notation in the rescaled metric $g'_{mn}$, equations (\ref{bisp1}, \ref{bisp2}) 
can be written succinctly as 
\beal
0&=d\psp+F\wedge\wps^-\nn\\
0&=e^{-3\Delta}d(e^{3\Delta}\psm)+\star F\wedge\wps^-
-f\wedge\psp+4m\wps^-
~,
\label{ole1}
\end{align}
where 
\beal
\wps^-:=\frac{1}{2}(\wps-\star\wps)=
\frac{1}{8}P_+\Big\{ \wps_m\ga^m+\frac{1}{5!}\wps_{m_1\dots m_5}\ga^{m_1\dots m_5} \Big\}~.
\label{p1}
\end{align}
We also define 
\beal
\wps^+:=\frac{1}{2}(\wps+\star\wps)=
\frac{1}{8}P_+\Big\{ \frac{1}{3!}\wps_{m_1\dots m_3}\ga^{m_1\dots m_3}
+\frac{1}{7!}\wps_{m_1\dots m_7}\ga^{m_1\dots m_7}\Big\}~.
\label{p2}
\end{align}
Multiplying (\ref{3}), (\ref{4}) on the left by $\xi^-\ga_{mp}$, $\xi^+\ga_{mp}$ 
respectively and adding, we obtain
\beal
\frac{1}{12}F_{[m}{}^{ijk}\psp_{p]ijk}+\frac{1}{2}f_i\wps^i{}_{mp}=-3\wps_{[m}\partial_{p]}\Delta
~.
\label{e5}
\end{align}
Multiplying (\ref{1}), (\ref{2}) on the left by $\xi^-\ga_{p}$, $\xi^+\ga_{p}$ 
respectively and adding, taking (\ref{e5}) into account, we obtain
\beal
0=\partial_{[m}(e^{3\Delta}\wps_{p]} )
~.
\label{bisp3}
\end{align}
Multiplying (\ref{3}), (\ref{4}) on the left by $\xi^-\ga_{mpqr}$, $\xi^+\ga_{mpqr}$ 
respectively and subtracting, we obtain
\beal
\frac{3}{4}\psm_{[mp}{}^{ij}F_{qr]ij}+\frac{1}{2}\wps_{mpqr}{}^if_i=
-6\wps_{[mpq}\partial_{r]}\Delta+\frac{1}{4}(\star F)_{mpqr}\psp+
\frac{1}{4}F_{mpqr}\psm
+3m\psp_{mpqr}
~.
\label{e7}
\end{align}
Multiplying (\ref{1}), (\ref{2}) on the left by $\xi^-\ga_{pqr}$, $\xi^+\ga_{pqr}$ 
respectively and adding, taking (\ref{e7}) into account as well as the identity
\beal
\frac{1}{24}\Phi^{\pm}_{mpqr}{}^{n_1\dots n_4}F_{n_1\dots n_4}=\pm\Phi^{\pm}(\star F)_{mpqr}~,
\end{align}
we obtain
\beal
0=e^{-6\Delta}\partial_{[m}(e^{6\Delta}\wps_{pqr]} )
+\frac{1}{4}(\star F)_{mpqr}\psp-\frac{1}{4}F_{mpqr}\psm
+m\psp_{mpqr}
~.
\label{bisp4}
\end{align}
Multiplying (\ref{3}), (\ref{4}) on the left by $\xi^-\ga_{mpqrst}$, $\xi^+\ga_{mpqrst}$ 
respectively and subtracting, we obtain
\beal
5\psp_{[mpq}{}^{i}F_{rst]i}+\frac{1}{2}\wps_{mpqrst}{}^if_i=
-9\wps_{[mpqrs}\partial_{t]}\Delta+\frac{1}{4}\psp_{[mpqrs}{}^{ijk}F_{t]ijk}
~.
\label{e9}
\end{align}
Multiplying (\ref{1}), (\ref{2}) on the left by $\xi^-\ga_{pqrst}$, $\xi^+\ga_{pqrst}$ 
respectively and adding, taking (\ref{e9}) into account as well as the identity
\beal
0=\Psi^{\pm}_{[mpqrs}{}^{ijk}F_{t]ijk}~,
\end{align}
we obtain
\beal
0=\partial_{[m}(e^{9\Delta}\wps_{pqrst]} )
~.
\label{bisp5}
\end{align}
Multiplying (\ref{3}), (\ref{4}) on the left by $\xi^-\ga_{mpqrstuv}$, $\xi^+\ga_{mpqrstuv}$ 
respectively and subtracting, we obtain
\beal
12\wps_{[mpqrstu}\partial_{v]}\Delta=\frac{35}{2}\psm_{[mpqr}F_{stuv]}
+3m\psp_{mpqrstuv}
~.
\label{e11}
\end{align}
Multiplying (\ref{1}), (\ref{2}) on the left by $\xi^-\ga_{pqrstuv}$, $\xi^+\ga_{pqrstuv}$ 
respectively and subtracting, taking (\ref{e11}) into account, 
we obtain
\beal
0=\partial_{[m}(e^{12\Delta}\wps_{pqrstuv]} )+m\psp_{mpqrstuv}
~.
\label{bisp6}
\end{align}
Passing to the bispinor notation for the rescaled metric, taking into account the identity
\beal
(\star F)_{[mpqr}\psp_{stuv]}-F_{[mpqr}\psm_{stuv]}=0~,
\end{align}
equations (\ref{bisp3}, \ref{bisp4}, \ref{bisp5}, \ref{bisp6}) can be written succinctly as
\beal
0&=e^{-3\Delta/2}d(e^{3\Delta/2}\wps^-)\nn\\
0&=e^{-3\Delta/2}d(e^{3\Delta/2}\wps^+)+2(\star F\wedge \psp-F\wedge\psm)
+8m\psp
~.
\label{ole2}
\end{align}
It is easy to see that the integrability of (\ref{ole1}, \ref{ole2}) follows from the 
equations-of-motion and Bianchi identities. 
The `asymmetry' between $\Psi^{\pm}$ in equation (\ref{ole1}) disappears in the 
massless limit 
for constant 
warp factor and in the absence of fluxes along the noncompact spacetime directions; i.e. 
for $\Delta=$constant, $m$, $f=0$. In this case we have:
\beal
0&=d\psp+F\wedge\wps^-\nn\\
0&=d\psm+\star F\wedge\wps^-\nn\\
0&=d\wps^-\nn\\
0&=d\wps^++2(\star F\wedge \psp-F\wedge\psm)
\end{align}
The integrability of the above equations follows from the 
equations-of-motion and Bianchi identities, which now read
$dF$, $d\star F=0$. 
Note, however, that setting $m=0$ excludes any solutions with nonzero fluxes 
if $X$ is smooth and compact, as was already noted in section \ref{nsusy}.

%
%

\end{document}